\documentclass[lettersize,journal]{IEEEtran}
\usepackage{amsmath,amsfonts}
\usepackage{algorithm}
\usepackage{array}
\usepackage{url}
\usepackage[hidelinks]{hyperref}
\usepackage{textcomp}
\usepackage{stfloats}
\usepackage{multirow}
\usepackage{verbatim}
\usepackage{graphicx}
\usepackage{cite}
\usepackage{xcolor}
\usepackage[caption=false,font=footnotesize]{subfig}

\begin{document}

\title{On Multi-entity, Multivariate Quickest Change Point Detection}

\author{
%\thanks{This paper was produced by the IEEE Publication Technology Group. They are in Piscataway, NJ.}% <-this % stops a space

\IEEEauthorblockN{Bahar Kor\IEEEauthorrefmark{1}, Bipin Gaikwad\IEEEauthorrefmark{2}, Abani Patra\IEEEauthorrefmark{3}, Eric L.~Miller\IEEEauthorrefmark{4}}
\thanks{\IEEEauthorrefmark{1}Bahar Kor is with the Department of Electrical and Computer Engineering, Tufts University, Medford, MA, USA (email: bahar.kor@tufts.edu). She is the corresponding author.}
\thanks{\IEEEauthorrefmark{2}Bipin Gaikwad is with the Department of Electrical and Computer Engineering, Tufts University, Medford, MA, USA (email: bipin.gaikwad@tufts.edu).}
\thanks{\IEEEauthorrefmark{3}Abani Patra is with the Department of Computer Science, Tufts University, Medford, MA, USA (email: abani.patra@tufts.edu).}
\thanks{\IEEEauthorrefmark{4}Eric L.~Miller is with the Department of Electrical and Computer Engineering, Tufts University, Medford, MA, USA (email: eric.miller@tufts.edu).}
}
%{Kor \MakeLowercase{\textit{et al.}}: On Multi-entity, Multivariate Quickest Change Point Detection}

\markboth{}%
{Kor \MakeLowercase{\textit{et al.}}: On Multi-entity, Multivariate Quickest Change Point Detection}


\maketitle

\begin{abstract}
We propose a framework for online Change Point Detection (CPD) from multi-entity, multivariate time series data, motivated by applications in crowd monitoring where traditional sensing methods (e.g., video surveillance) may be infeasible. Our approach addresses the challenge of detecting system-wide behavioral shifts in complex, dynamic environments where the number and behavior of individual entities may be uncertain or evolve. We introduce the concept of Individual Deviation from Normality (IDfN), computed via a reconstruction-error-based autoencoder trained on normal behavior. We aggregate these individual deviations using mean, variance, and Kernel Density Estimates (KDE) to yield a System-Wide Anomaly Score (SWAS). To detect persistent or abrupt changes, we apply statistical deviation metrics and the Cumulative Sum (CUSUM) technique to these scores. Our unsupervised approach eliminates the need for labeled data or feature extraction, enabling real-time operation on streaming input. Evaluations on both synthetic datasets and crowd simulations, explicitly designed for anomaly detection in group behaviors, demonstrate that our method accurately detects significant system-level changes, offering a scalable and privacy-preserving solution for monitoring complex multi-agent systems.

In addition to this methodological contribution, we introduce new, challenging multi-entity multivariate time series datasets generated from crowd simulations in Unity and coupled nonlinear oscillators. To the best of our knowledge, there is currently no publicly available dataset of this type designed explicitly to evaluate CPD in complex collective and interactive systems, highlighting an essential gap that our work addresses.
\end{abstract}


\section{Introduction}
\label{sec: introduction}

Change point detection refers to the process of identifying significant changes in statistical properties of a system, often signaling a transition between states or behaviors. These shifts, or ``change points", can indicate underlying issues such as system failures, unexpected disturbances, or critical transitions in complex systems \cite{introcpd}.

The motivation for this work comes from the monitoring of crowd behavior, particularly in safety and security applications where traditional video surveillance may be infeasible. Factors such as privacy concerns, infrastructure limitations, high computational costs, and challenging environmental conditions can render video-based approaches impractical \cite{constraints}. To address this, we investigate how sensor-based data streams provided by agents in these venues, such as accelerometer time series, can serve as an effective alternative for promptly detecting major incidents.\footnote{In terms of how such data would be acquired, we envision a system in which individuals such as students, teachers, or parents in a school, commuters at a transit station, or attendees at a large event \cite{crowdtaxonomy} opt into providing such data to a central processing facility \cite{MCS}. Although the social and technical details of such a system are both interesting and highly non-trivial, the focus of this work is on the value of the data to identify emergencies, such as stampedes, security threats, or other abnormal events.} 
By analyzing patterns in collective behavior, CPD can detect sudden changes in crowd dynamics and facilitate identification of sudden shifts, enabling real-time interventions that mitigate risks to public safety. 

Detecting these changes promptly is crucial for effective event management, such as in transport hubs and urban planning, to ensure public safety \cite{CoenoSense}.

Although crowd safety is our primary application, CPD has been extensively studied in various domains. In autonomous transportation, CPD has been applied to detect behavioral anomalies in connected vehicle networks \cite{auto-cpd}. In industrial systems, detecting sudden changes in sensor readings is critical for the early identification of system failures and enabling timely corrective actions \cite{industrial-CPD-2}. In finance, CP analysis is used to establish investment strategies by identifying regime changes in financial markets \cite{finance_CPD}. These and other applications highlight that CPD is a broad problem that arises in various contexts involving different types of systems, data structures, and goals. 

Our specific interest in this work lies in the \textit{online detection} of a single change point, which often corresponds to a system failure or a critical event. This approach is commonly referred to as the quickest Change Point Detection (qCPD), where the objective is to identify the change as quickly as possible after it occurs \cite{qCPD}. In our setting, the data provided to the system is \textit{multi-entity multivariate time series}. Examples include movement patterns of individuals in a crowd, sensor readings from multiple machines in an industrial network, or time series generated by financial agents in a trading system. These settings share the challenge of detecting abrupt transitions in complex, high-dimensional environments, making qCPD a crucial tool for making timely decisions and interventions.
\IEEEpubidadjcol
\subsection{Challenges in Multi-Entity qCPD}

Although researchers have extensively studied qCPD in the context of multivariate time series, extending these techniques to multi-entity systems introduces additional layers of complexity. 
Traditional CPD methods, such as Bayesian CPD \cite{BCPD}, kernel-based approaches \cite{kernelCPD}, and CUSUM \cite{cusum}, typically operate on multivariate time series. For $F$ signals sampled $T$ times, such datasets can be thought of as a $F \times T$ matrix. 

These methods often assume that the statistical properties of the data can be characterized by predefined distributions, kernels, or likelihood models \cite{MultiAgent}. However, data generated by multi-entity systems has a more complex structure, namely a $P \times F \times T$ tensor where $P$ represents the number of entities. In such systems, changes often emerge not from isolated deviations in individual entities but from collective patterns. As a result, standard CPD techniques often fail to account for the complexities introduced by interactions among multiple entities. Naively flattening the data into an $PF \times T$ matrix or applying a set of independent single-entity CPD overlooks interactions across entities \cite{MultiAgent}. Moreover, in many real-world settings, the number of active entities is not known a priori. It can change dynamically, making it even more challenging to model the system's behavior accurately.

To address these limitations, we use an autoencoder-based approach to learn a deterministic model of individual entity dynamics. Rather than relying on probabilistic or statistical techniques, we detect change points by analyzing patterns in the reconstruction error. This method enables system-wide change point detection by identifying collective deviation patterns without relying on predefined probabilistic assumptions. Additionally, it remains effective even in scenarios where the number of entities is uncertain or changes dynamically.

This journal paper extends our previous work \cite{ours} by framing the problem as change point detection in multi-entity systems. Unlike the previous work, which used only the mean of per-entity reconstruction errors, this work introduces new aggregation methods, including variance and Wasserstein distance with a KDE, and applies a CUSUM-based procedure for online system-level CPD. The approach is evaluated on new, challenging datasets, including coupled Chen chaotic oscillators and an additional Unity crowd simulation scenario, which we plan to make publicly available, and includes comparisons with an additional state-of-the-art model.

\subsection{Key Contributions}

This paper presents a novel framework for qCPD in multi-entity systems using an autoencoder-based approach.  

\begin{figure}[t]
\centerline{\includegraphics[width=0.5\textwidth]{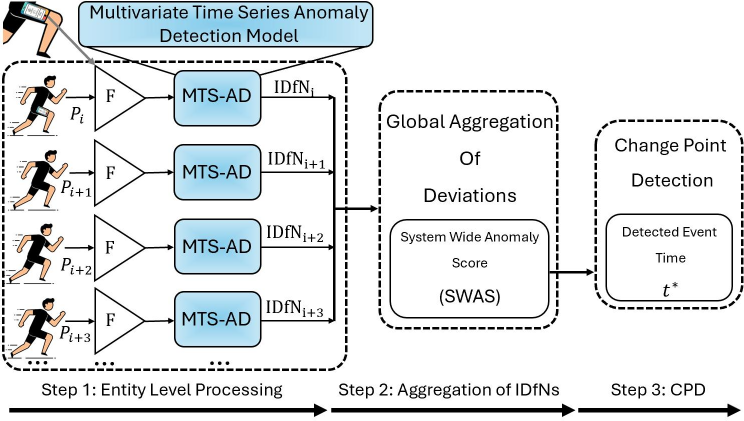}}
\caption{\textbf{Overview scheme of framework.} A high-level representation of the proposed framework, illustrating the key components and workflow.}
\label{fig:framework}
\end{figure}

Our key contributions are as follows.
\begin{itemize}
    \item We propose a method for detecting a change point in multi-entity systems by modeling the collective behavior of interacting entities using autoencoder-based anomaly detection.
    \item We introduce \textit{Individual Deviation from Normality} (IDfN) metric to quantify each entity’s deviation from learned normal behavior. These deviations are then aggregated into a \textit{System-Wide Anomaly Score} (SWAS), capturing system-level behavioral changes.
    \item We combine SWAS with CPD to efficiently detect a change point in dynamic, high-dimensional environments.
    \item We present a framework for online qCPD that handles real-time data streams, suitable for practical applications.
    \item Our approach eliminates the need for feature extraction or labeled training data, utilizing unsupervised learning to perform scalable, low-cost anomaly detection (AD) without requiring explicit dimensionality reduction.
    \item We introduce and make available to the community new, challenging multi-entity multivariate time series datasets, generated from crowd simulations in Unity and coupled nonlinear oscillators, designed specifically to benchmark CPD and AD methods in complex interactive systems.
\end{itemize}

The remainder of the paper is organized as follows. In Section \ref{sec: related_works}, we review existing methods for CPD, highlighting their limitations for multi-entity multivariate systems. Section \ref{sec: prob_formulation} formally defines the problem and outlines the proposed method. In Section \ref{sec: proposed_method}, we describe the details of our approach, including the anomaly detection model and the CPD framework. In Section \ref{sec: dataset}, we introduce the dataset used for evaluation. Section \ref{sec: experiment} details the experimental setup, while Section \ref{sec: result} discusses the results and performance analysis. Finally, Section \ref{sec: conclusion} summarizes our findings and suggests future directions for future work.

\section{Related Works}
\label{sec: related_works}

CPD in multi-entity and high-dimensional systems has become increasingly important, especially in applications such as crowd monitoring and public safety. Of recent concern is the detection of abnormal behavior in human crowds, where deviations in movement or behavior often indicate critical events such as panic, stampedes, or other emergencies.

Early work in this area primarily focused on vision-based approaches, leveraging surveillance footage to detect anomalies. For instance, the method proposed in \cite{CrowdChange_hist} identifies both global and local motion changes by analyzing 2D motion histograms over time and identifying clusters with similar spatial and velocity characteristics. Similarly, the study in \cite{hao-opticalflow} utilizes optical flow-based features to detect panic behavior, demonstrating the effectiveness of low-level motion features in identifying emergent crowd anomalies. Deep learning approaches have also been explored, such as in DeepROD \cite{deeprod}, which introduces a deep neural architecture for the real-time and online detection of panic behavior from surveillance footage. Another approach \cite{ammar} integrates a continuous video surveillance system tailored to specific public spaces, combining scene understanding with real-time anomaly analysis. While vision-based methods offer fine-grained spatial information, they also face several limitations, as outlined earlier in Section \ref{sec: introduction}, including privacy concerns, high computational demands, and sensitivity to environmental conditions.

\begin{figure*}[t]
\centering
\includegraphics[keepaspectratio, width=\textwidth]{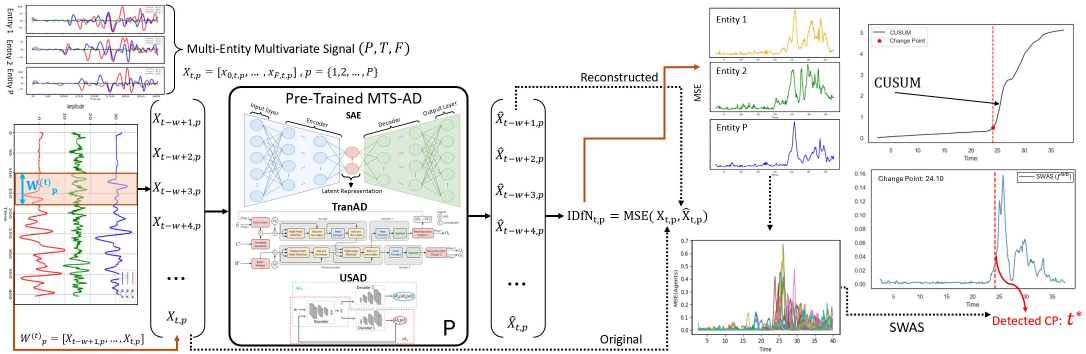}
\caption{\textbf{Proposed method pipeline.} The CPD framework processes sensor data streams from $P$ entities over $T$ time points using an entity-level pre-trained MTS-AD model. Each entity's $F$-dimensional sensor data is fed into this model, and the reconstruction error is computed using MSE and the IDfN score. The SWAS score is then obtained by aggregating the MSE values across $P$ entities at each time step $t$. The plots here show results for the crowd train simulation dataset described in Sec .~\ref{sec:crowdsim}, where we compute the SWAS using the \texttt{wass+KDE} method. A CPD algorithm is applied to detect change points, where the detected change point is shown as a red dot.} 
\label{fig:detailed_pipeline}
\end{figure*}

As a result, an emerging body of work focuses on sensor-driven and multi-modal approaches, which provide more scalable and privacy-preserving alternatives. For instance, \cite{geo-bio-CD} presents a system using geo-referenced biometric data from wearable and smartphones to detect and map panic behavior in real-time. Their method integrates physiological indicators (e.g., heart rate) with GPS data and applies machine learning to identify stress states and spatially localize incidents. This sensor-based approach provides a rich and low-intrusion data source, with metrics such as the Domino Effect Index (DEI), which enables the estimation of the severity of spreading panic. However, challenges remain in data privacy, as people in the crowd share their locations.

Another promising direction leverages Topological Data Analysis (TDA) to capture the underlying global structure in multi-entity systems. 
For instance, \cite{MA-CPD} introduces a CPD framework based on persistent homology to monitor changes in group dynamics by constructing Vietoris–Rips complexes from agent trajectories and extracting topological features such as Betti numbers and persistence. 
Although TDA-based methods are powerful, they are often computationally demanding and typically designed for offline analysis. Specifically, the construction of pairwise distance matrices and the persistence computation scale poorly with the number of entities, which can limit their use in real-time applications \cite{TDA-cost}.

Overall, vision-based methods offer fine-grained spatial understanding of crowd dynamics but face challenges in scalability, privacy, and real-time processing. In contrast, sensor-based approaches provide a lightweight, privacy-preserving alternative that is better suited for real-time deployment. Building on this direction, our work explores time-series data from individual entities, such as accelerometer readings, as a basis for scalable and responsive CPD. By leveraging individual-level analysis and aggregating deviations to capture system-level anomalies, we propose a practical framework that strikes a balance between expressiveness, computational efficiency, and feasibility for real-world deployment.
 
\section{Problem Overview and Formulation}
\label{sec: prob_formulation}
\subsection{Problem Overview}

Figure~\ref{fig:framework} illustrates the structure of our proposed framework for change point detection in multi-entity systems. It consists of three major components: 
 
\begin{itemize}
    \item \textbf{Individual Deviations from Normality (IDfN)} is an entity-level module that quantifies how each individual deviates from baseline behavior over time.
    
    \item \textbf{Global Aggregation of Deviations} integrates individual deviation scores into a system-level representation to capture the overall system dynamics. 
    
    \item \textbf{Change Point Detection} identifies the time at which the aggregated signal exhibits statistically significant shifts using CUSUM-based sequential detection.
\end{itemize}

In this setting, $P$ entities generate time series data, each associated with $F$ sensors that record the observation over $T$ time steps. Given training time series data with normal behavior, our objective is to detect a change, if it has occurred, with minimum detection delay in any unseen test time series that shares the same modality as the training data, ensuring that the change is detected as quickly as possible after it happens.

\subsection{Problem Formulation}

At any given time step $t \in [T] \equiv \{1,2,\dots, T\}$, the reading from sensor $s \in [F]$ for entity $p \in [P] \equiv \{1,2,\dots, P\}$ is denoted as datum $ x_{s,t,p}$. The complete sensor observation for entity $p$ at time $t$ forms a vector:
\begin{equation}
    X_{t,p}= [x_{1,t,p}, x_{2,t,p}, \dots, x_{F,t,p}] \in \mathbb{R}^{F}
\end{equation}

By organizing the data across all sensors, times, and entities, we structure the dataset as a third-order tensor $\mathcal{X} \in \mathbb{R}^{F \times T \times P}$.

The goal of CPD is to identify the time step $t^* \in [T]$ at which a significant change occurs in the behavior of a monitored system comprising $P$ entities. 

CPD algorithms are broadly categorized into two types:
\begin{itemize}
    \item \textbf{Offline methods} assume access to the entire time series and detect changes retrospectively. These are suitable for post hoc analysis and diagnostic applications.
    \item \textbf{Online methods} process data sequentially in real-time, detecting changes immediately after they occur \cite{introcpd}.
\end{itemize}
In this work, we develop an online CPD algorithm. At each time $t$, the algorithm computes a test statistic $C_t$ based on past observations. A change point is declared once the statistic exceeds the threshold $h$:

\begin{equation}
\hat{t} = \min \left\{ t : C_t > h \right\}, \quad 
\begin{cases}
\hat{t} \in [T], & \text{if a crossing occurs}, \\
\infty, & \text{otherwise}.
\end{cases}
\end{equation}

Here, $\hat{t}$ is the stopping time at which the algorithm signals a change, and the threshold $h$ controls the trade-off between rapid detection and false alarms.

The performance of a CPD algorithm in the online setting is commonly evaluated using the detection delay, defined as:
\begin{equation}
    \text{Detection delay: } \delta = \hat{t} - t^* \quad \text{subject to } \hat{t} \geq t^*.
\end{equation}

This metric captures how quickly the method responds after the actual change $(t^*)$ occurs, while maintaining a low false alarm rate, particularly when no change occurs during the observation window.

In the following sections, we describe our proposed method for learning normal system patterns, calculating change scores, and applying CPD algorithms to detect critical transitions in complex, high-dimensional, multi-entity time series data.

\begin{figure}[t]
\centerline{\includegraphics[width=0.5\textwidth]{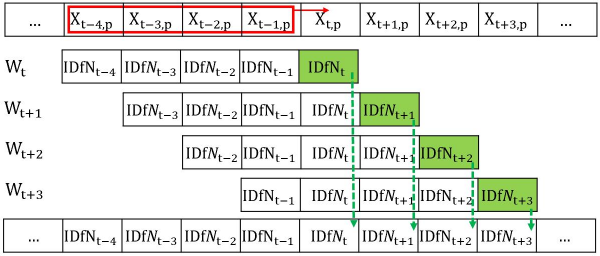}}
\caption{\textbf{Sliding window-based scoring using IDfN.} At each time step, a fixed-length window $w$ is applied over the input time series. IDfN scores are computed for each time step within the window based on reconstruction error, and only the IDfN score associated with the current window time step is retained for real-time applicability.}
\label{fig:sliding_window}
\end{figure}

\section{Proposed Method}
\label{sec: proposed_method}

In this section, we present our proposed method for qCPD in multi-entity, multivariate time series data. 
Referring to Figure~\ref{fig:detailed_pipeline}, we consider a multi-entity, multivariate time series dataset where each entity $p \in [P]$ reports $F$-dimensional observations over $T$ time steps. We first train the model on normal data collected from all entities to learn typical system behavior. During testing, we process each entity’s data independently, and deviations from the learned representation are quantified using IDfNs. To assess system-level behavior, we aggregate IDfN scores across all entities at each time step to compute SWAS. This scalar time series provides a holistic representation of the system behavior capable of capturing collective deviations from normality. Using SWAS, we detect the time step $\hat{t}$ when the system undergoes a significant change, marking a potential change point. 

\subsection{Preprocessing}

To ensure meaningful comparisons across different sensors and time steps, we apply \textbf{z-score normalization} to the data:

\begin{equation}  
    x_{s,t,p} \leftarrow \frac{x_{s,t,p} - \mu_s}{\sigma_s}
\end{equation}

Where $\mu_s$ and $\sigma_s$ are the mean and standard deviation of the training data sensor readings across all time steps and entities for sensor $s$:
\begin{equation}
    \mu_s = \frac{1}{PT} \sum_{p=1}^{P} \sum_{t=1}^{T} x_{s,t,p}, \quad
    \sigma_s = \sqrt{\frac{1}{PT} \sum_{p=1}^{P} \sum_{t=1}^{T} (x_{s,t,p} - \mu_s)^2}
\end{equation}

These normalization parameters are applied consistently across all entities and time steps. This ensures that the data from each sensor is on a comparable scale and eliminates the influence of differing units or ranges across sensors.

To effectively capture temporal dependencies, we segment the time series into overlapping windows of length $w$ with a stride of $1$. For entity $p$ at time $t$, the window is defined as:
\begin{equation}
    W^{(t)}_p = [X_{t-w+1,p}, X_{t-w+2,p}, \dots, X_{t,p} ] \in \mathbb{R}^{F \times w}.
\end{equation}
For time steps $t<w$, replication padding \cite{tranAD} is applied to extend the initial observations, ensuring that the temporal window is consistent across all time steps.

To aggregate data from all entities $p \in [P]$ at time $t$, we combine sensor readings from all $F$ sensors within the $t$-th window into a tensor $\mathcal{S}_t\in \mathbb{R}^{F \times w \times P}$. Each frontal face \cite{tensor-decom} of this tensor corresponds to $W^{(t)}_p$.

\begin{figure}[t]
\centerline{\includegraphics[width=0.5\textwidth]{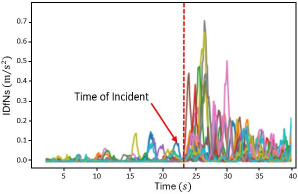}}
\caption{\textbf{Raw IDFN score over time computed for $P$ entities in the crowd-train simulation dataset presented in Sec.~\ref{sec:crowdsim}.} The ground truth time of the incident (vertical red dashed line) coincides with a clear rise in the score, illustrating its sensitivity to behavioral changes in the data.}
\label{fig:idfn}
\end{figure}

\subsection{Entity-Level Processing}

The first step detects deviations from normal behavior at the individual entity level. To achieve this, we employ a pre-trained Multivariate Time Series Anomaly Detection (MTS-AD) model, which reconstructs each entity’s time series and quantifies deviations. In Sec.~\ref{sec:mtsad} we discuss the MTS-AD models used in this paper. Higher IDfN scores indicate greater deviations from the learned normal patterns, suggesting potential behavioral anomalies at the entity level. The model reconstructs each entity’s time series and computes the reconstruction error at each time step.

The deviation is quantified using the Mean Squared Error (MSE), referred to as the IDfN score:
\begin{equation}
    \text{IDfN}_{t,p}= \frac{1}{F} \sum_{s=1}^{F} \left( \hat{x}_{s,t,p} - x_{s,t,p} \right)^2
\label{IDfN}
\end{equation}

In \eqref{IDfN}, $x_{s,t,p}$ and $\hat{x}_{s,t,p}$ are the original and reconstructed values of the sensor $s$ at time step $t$ for the entity $p$, respectively. The score reflects how much an entity’s behavior deviates from the learned norm. IDfN scores are computed at each time step within the sliding window $W^{(t)}_p$. For real-time processing and efficient change detection, only the score from the current window time step is retained (see Fig.~\ref{fig:sliding_window}) \cite{lastPoint}. Fig.~\ref{fig:idfn} shows IDfN evolution over time for all $P$ entities, with the ground truth incident time marked by a red dashed line.

\subsection{Global Aggregation of IDfNs}

To assess the collective system behavior over time, we aggregate the IDfN scores computed at the entity level. This aggregation step captures the overall system dynamics by summarizing how each entity deviates from its baseline behavior. We explore three ways of summarizing the data, all of which are described below. The first two, mean and variance, are standard statistical measures. The third treats the collection of IDfN data at each time as samples from a time-varying probability density function. It uses kernel density estimation (KDE) \cite{KDE} methods to estimate this function. 

\subsubsection{IDfN Mean} The first method computes the mean of the IDfN scores across all $P$ entities at each time step $t$, providing a central tendency of the system's behavior. The aggregated score at time $t$ is given by:
\begin{equation}
        S^{\mu}_t = \frac{1}{P} \sum_{p=1}^{P} \text{IDfN}_{t,p}
    \label{anomaly_score_mean}
\end{equation}

\subsubsection{IDfN Variance}
The second method aggregates IDfNs by calculating the variance of IDfN scores between all $P$ entities at each time step $t$. This approach measures the dispersion in the behavior of entities and is defined as:
\begin{equation}
    S^{\sigma^2}_t = \frac{1}{P} \sum_{p=1}^{P} (\text{IDfN}_{t,p} - S^{\mu}_t)^2
\label{anomaly_score}
\end{equation}
where $S^{\mu}_t$ is the mean of the IDfNs at time $t$.

\begin{figure}[t]
\centerline{\includegraphics[width=0.5\textwidth]{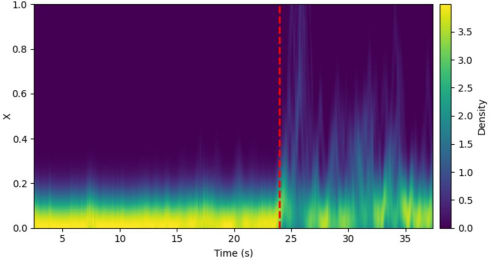}}
\caption{KDE heatmap showing the temporal evolution of the estimated density function $\hat{f}_t(x)$ of IDfN scores in the train simulation dataset. The plot reveals shifts in the distribution over time that indicate anomalies, with a notable change at the ground truth incident time $(24s)$, shown as a red dashed line.}
\label{fig:kde}
\end{figure}

\subsubsection{IDfN Distribution}
To model the distribution of deviations at the entity level, the third aggregation method uses KDE on IDfN scores at each time step $t$. KDE provides a smooth, non-parametric estimate of the probability density function of the reconstruction errors, enabling sensitivity to subtle distributional shifts of IDfN scores among entities. Temporal variations in $\hat{f}_t(x)$ reflect evolving system dynamics and may indicate global behavioral shifts, as shown in Fig.~\ref{fig:kde}.

The estimated density function $\hat{f}_t(x)$ at time step $t$ is defined as:
\begin{equation}
        \hat{f}_t(x) = \frac{1}{Ph} \sum_{p=1}^{P} K\left( \frac{x - \text{IDfN}_{t,p}}{h} \right)
\end{equation}
where $h$ is the bandwidth that controls the smoothness of the estimate, and $K(\cdot)$ is the kernel function. In this work, we use the Gaussian kernel function:
\begin{equation}
        K(u) = \frac{1}{\sqrt{2\pi}} \exp\left( -\frac{u^2}{2} \right)
\end{equation}
and determine the bandwidth via grid search with 5-fold cross-validation on the training set as described in \cite{CPD-Bandwidth}.

In summary, the three aggregation methods—mean, variance, and KDE—provide complementary statistical characterizations of system behavior at each time step. 

\subsection{Change Point Detection}

\subsubsection{Statistical Deviation Metrics}
\label{sec: statistical_metrics}
This section describes how we detect changes in system behavior by comparing the current distribution of IDfN scores with a baseline distribution derived from training data. We use different metrics depending on the aggregation method used for IDfNs:

\paragraph{Mean-Based Deviation Metric} 
For mean aggregation, we compute the absolute deviation between the current aggregated score at time $t$ and the reference mean obtained from the training data. The reference mean $\mu_{\text{train}}$ is calculated as:
\begin{equation}
    \mu_{\text{train}} = \frac{1}{PT} \sum_{t=1}^{T} \sum_{p=1}^{P} \text{IDfN}^{train}_{t,p}
\label{eq:mu_train}
\end{equation}
The deviation score at each time step $t$ is then defined as:
\begin{equation}
    f^{\mu}(t) = | S^{\mu}_t - \mu_{\text{train}} |
\label{eq:mu_score}
\end{equation}

\paragraph{Variance-Based Deviation Metric}
For variance aggregation, we measure deviation by calculating the absolute difference between the current variance of the IDfN scores $S^{\sigma^2}_t$ and the reference variance obtained from the training data. The reference variance, $\sigma^2_{train}$, is defined as:
\begin{equation}
    \sigma^2_{\text{train}} = \frac{1}{PT} \sum_{t=1}^{T} \sum_{p=1}^{P} \left( \text{IDfN}^{train}_{t,p} - \mu_{\text{train}} \right)^2
\label{eq:sigma_train}
\end{equation}

The variance deviation score at time $t$ is then computed as:

\begin{equation}
    f^{\sigma^2}(t) = \left| S^{\sigma^2}_t - \sigma^2_{\text{train}} \right|
\label{eq:sigma_score}
\end{equation}

\begin{table*}[t]
\begin{center}
\caption{STATISTICAL SUMMARY OF DATASETS
}
\label{tab1}
\begin{tabular}{|l|c|c|c|c|c|c|c|c|}
\hline
\textbf{Dataset Name} & \textbf{\#P} & \textbf{\#F} & \textbf{\#Normal Samples} & \textbf{\#T} & 
\textbf{\#Abnormal Samples} & \textbf{\#T} & \textbf{\#Sampling Rate} & \textbf{\#GT (s)} \\
\hline
Auto-regressive (Mean) & 10 & 1 & 1200 & 300 & 1000 & 300 & 1 & 150 \\
\hline
Auto-regressive (Variance) & 10 & 1 & 1200 & 300 & 1000 & 300 & 1 & 150 \\
\hline
Coupled Chen & 4 & 3 & 500 & 1900 & 500 & 1900 & 0.01 & 9 \\
\hline
Crowd (Collision) & 30 & 3 & 400 & 465 & 300 & 350 & 0.1 & 23 \\
\hline
Crowd (Train Station) & 20 & 3 & 1000 & 650 & 500 & 400 & 0.1 & 24 \\ 
\hline
\end{tabular}
\end{center}
\end{table*}

\paragraph{Wasserstein Distance-Based Metric}

We measure the distributional shift between $\hat{f}_t$ and a reference distribution $\hat{f}_{\text{train}}$. To obtain the reference, the normal training data is passed through the pre-trained model to obtain reconstruction errors for all entities across all training time steps. These reconstruction errors, $\{ \text{IDfN}^{train}_{t,p} \}_{t=1,p=1}^{T, P}$, are used to fit a KDE, resulting in $\hat{f}_{train}$ \cite{KDE_train}:
\begin{equation}
    \hat{f}_{\text{train}}(x) = \frac{1}{(PT)h} \sum_{t=1}^{T} \sum_{p=1}^{P} K\left( \frac{x - \text{IDfN}_{t,p}}{h} \right),
\end{equation}

We use the Wasserstein-1 distance $(WD)$, also known as the Earth Mover's Distance, to quantify how much the distribution of entity-level deviations at time $t$ has drifted from normal system behavior: 
\begin{equation}
    f^{\text{WD}}(t) = WD\big(\hat{f}_t, \hat{f}_{\text{train}}\big),
\label{eq:wd_score}
\end{equation}

For a rigorous definition and mathematical properties of the Wasserstein-1 distance, we refer the reader to Part~1.6 of \cite{wass-dist}.

The deviation scores are then input to the CUSUM algorithm to detect the change point.
\subsubsection{CUSUM}
\label{sec: cusum}
To detect persistent or cumulative changes in the system, we apply the CUSUM technique to the computed scores in Section~\ref{sec: statistical_metrics}. For each aggregation method, we apply CUSUM on the corresponding change score $f^{\alpha}(t)$, for $\alpha \in \{ \mu, \sigma^2, \text{WD}\}$. The CUSUM is updated iteratively as:
\begin{equation}
C_t =
\begin{cases}
    0, & \text{if } t = 0 \\
    C_{t-1} + f^{\alpha}(t), & \text{if } t \geq 1
\end{cases}
\label{eq:cusum}
\end{equation}

Here, $C_t$ represents the cumulative evidence of a change up to time $t$, allowing us to detect abrupt and gradual deviations. If $C_t$ exceeds the threshold, it indicates the presence of a significant change at time $t$.

\subsubsection{Adaptive Thresholding}

To improve robustness and adaptivity in change point detection, we apply KDE-based thresholding on the CUSUM scores introduced in Section~\ref{sec: cusum}, following the sequential density-based anomaly detection approach of \cite{estimate}. 

To stabilize the variance and ensure numerical robustness, we apply a log transformation \cite{log-transform}.

\begin{equation}
    \tilde{C}_i = \log(1 + C_i), \quad \text{for } i=1, \dots, t-1
\end{equation}

Using the transformed values, we use KDE to model the distribution of values up to time $t-1$:
\begin{equation}
    \hat{p}_{t-1}(x) = \frac{1}{(t-1)h} \sum_{i=1}^{t-1} K\left( \frac{x - \tilde{C}_i}{h} \right).
\end{equation}

At the current time $t$, the log-transformed CUSUM value $\tilde{C}_t$ is evaluated under the estimated density function, $\hat{p}_{t-1}(\tilde{C}_t)$, and a change is declared at time $t$ if:
\begin{equation}
    \hat{p}_{t-1}(\tilde{C}_t) < \delta \;\;\Rightarrow\;\; \hat{t} = t
\end{equation}

The threshold $\delta$ is selected as the $\alpha$-quantile (e.g., 5th percentile) of the density estimates computed on the training\cite{threshold} CUSUM scores. Since KDE requires a sufficient number of past values to produce a reliable estimate, we initialize the KDE using the first $2s$ CUSUM values time window. Change point evaluation begins only after this initial window.

\begin{figure*}[t]
  \centering
  \subfloat[Normal Case (No Change Point).\label{fig:chen_normal}]{
    \includegraphics[width=0.48\textwidth]{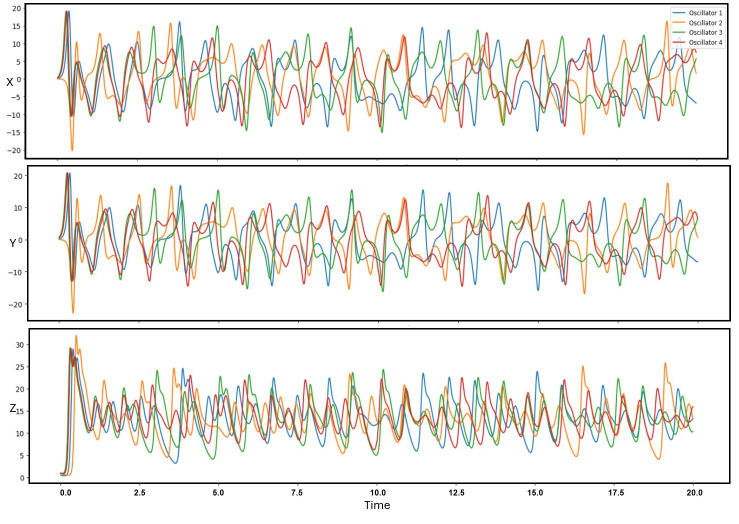}
  }
  \hfill
  \subfloat[Abnormal Case (Change at \( t = 10s \)).\label{fig:chen_change}]{
    \includegraphics[width=0.48\textwidth]{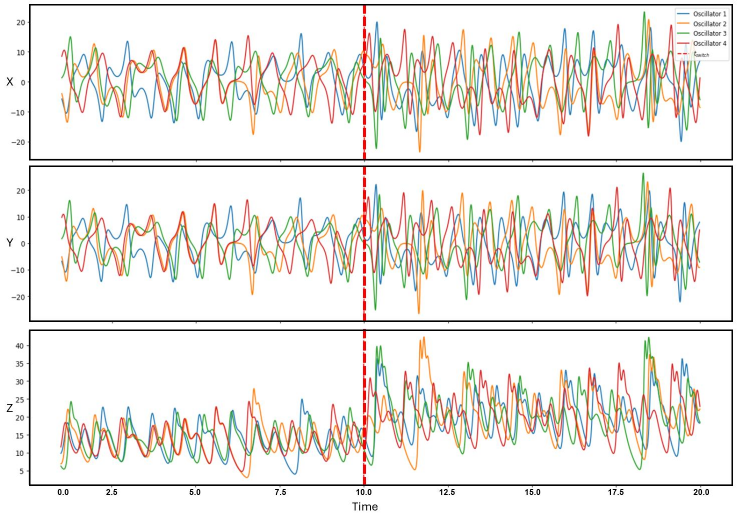}
  }
  \caption{Example time series from the synthetic chaotic oscillator dataset. (a) A stable, coupled system without any structural change. (b) A system with a change point at $t = 10s$, where the coupling is disabled and parameter $c$ is modified, inducing a regime shift in the dynamics. In both cases, the first 100 time steps $1s$ are discarded to eliminate transient behavior and focus the analysis on steady-state dynamics.}
\label{fig:chen_img}
\end{figure*}

\begin{figure*}
  \centering
  \subfloat[Train station simulation: Normal (left) vs Abnormal (right).\label{fig:train_scenario}]{
    \includegraphics[width=0.48\textwidth]{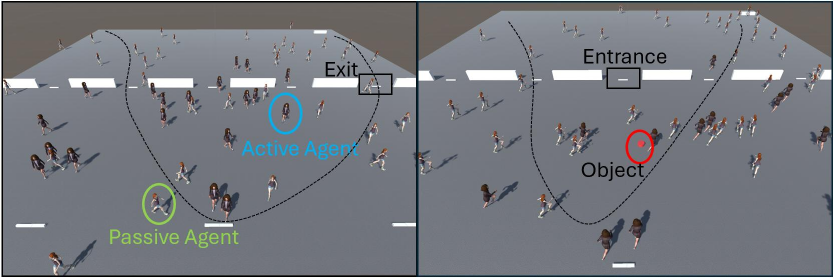}
  }
  \hfill
  \subfloat[Crowd collision simulation: Normal (left) vs Abnormal (right).\label{fig:collision_scenario}]{
    \includegraphics[width=0.48\textwidth]{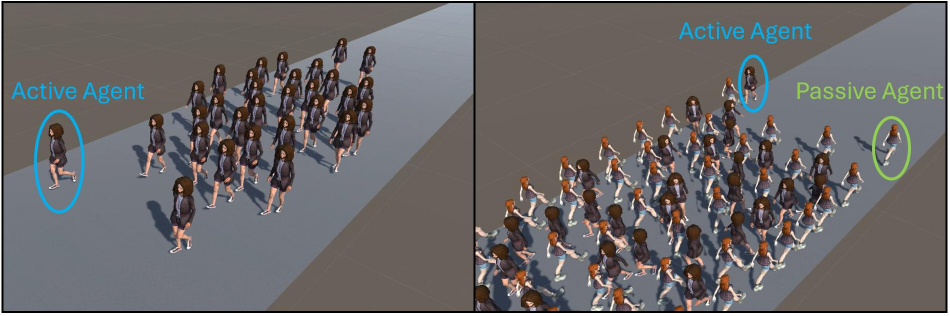}
  }
  \caption{Synthetic crowd scenarios: (a) Train station with emergency evacuation, (b) Group collision event.}
  \label{fig:all_scenarios}
\end{figure*}

\section{Datasets}
\label{sec: dataset}

Due to the lack of publicly available datasets suitable for CPD in multi-entity, multivariate time series data, we evaluate our method using synthetic data. We consider (i) an auto-regressive time series adopted from \cite{synthetic-dataset}, originally introduced for real-time change point detection in smart home sensor data, and (ii) a coupled chaotic system. In both cases, we modify the generation parameters to match our target scenario, introducing a single change point while preserving essential statistical properties.

To complement these datasets and better capture the dynamics of real-world environments, we develop a custom simulation environment in Unity. This environment supports the generation of multi-entity sensor streams under various crowd scenarios. Specifically, we design two distinct simulation scenarios: (i) a crowded train station, and (ii) a bidirectional corridor with pedestrian conflict. These scenarios enable a comprehensive evaluation of our method in varying densities, movement patterns, and interaction complexities.

All data and detailed documentation on the simulation setup and parameters, as well as post-processing scripts, are available in the accompanying GitHub repository: \url{https://github.com/bkor-git/AD-Project}.

%\subsection{Synthetic Dataset}
\subsection{Auto-Regressive (AR)}

We generated synthetic univariate time series per entity following the methodology from ~\cite{synthetic-dataset}, each with 300 time steps with a single change point at time step 150. This process is replicated for 10 entities to simulate a multi-entity setting. The datasets are designed to capture common statistical changes in sensor data, such as shifts in mean and variance, thereby providing a controlled and interpretable benchmark for evaluating change point detection methods.

\subsubsection{Dataset 1 (AR - Jump in Mean)} The data is generated using a first-order auto-regressive model defined as:
\begin{equation}
    y_{t,p} = \begin{cases}
        0, & t = 1, 2 \\
        0.6 y_{t-1,p} - 0.5 y_{t-2,p} + \epsilon_t, & t > 2
    \end{cases}
\end{equation}

where $\epsilon_t \sim \mathcal{N}(\mu, \sigma)$ is Gaussian noise. The mean of the noise changes from $\mu=0$ (for $t<150$) to $\mu=2$ (for $t \geq 150$), while the standard deviation remains fixed at $\sigma=0.5$. This introduces a distinct mean shift at the change point.

\subsubsection{Dataset 2 (AR - Jump in Variance)} Using the same AR structure, we introduce a change in the noise variance term instead of the mean. The noise is sampled from a zero-mean Gaussian distribution with a standard deviation of $\sigma = 0.1$ for $t < 150$ and $\sigma = 0.3$ for $t \geq 150$. This induces a noticeable increase in variability at the change point while keeping the mean constant, allowing the evaluation of sensitivity to variance change across entities.

\subsection{Coupled Chaotic Oscillator}

We construct a multi-entity multivariate time series dataset based on a network of $P=4$ identical Chen chaotic oscillators~\cite{chen}, coupled through a shared environmental variable inspired by~\cite{coupling}. This dataset is designed to benchmark change point detection methods in systems characterized by nonlinear, high-dimensional dynamics.

Each oscillator is described by three state variables $x_p(t)$, $y_p(t)$, and $z_p(t)$, where $p=1,2,\ldots,P$. The shared environment is represented by a scalar variable $w(t)$. The system evolves over time according to the following equations:

\begin{equation}
\begin{aligned}
\dot{x}_p(t) &= a \bigl( y_p(t) - x_p(t) \bigr) + \varepsilon_2 w(t), \\
\dot{y}_p(t) &= (c - a) x_p(t) - x_p(t) z_p(t) + c y_p(t), \\
\dot{z}_p(t) &= x_p(t) y_p(t) - b z_p(t).
\end{aligned}
\end{equation}

\begin{equation}
\dot{w}(t) = -\lambda_{\mathrm{env}} w(t) + \frac{\varepsilon_1}{P} \sum_{p=1}^{P} x_p(t).
\end{equation}

The parameters are set as $a=35.0$, $b=2.8$, and $\lambda_{\mathrm{env}}=1.0$. The coupling strengths $\varepsilon_1$ and $\varepsilon_2$ control the bidirectional interaction between the oscillators and the environment. Specifically, $\varepsilon_2$ governs the influence of the environment on the oscillators, while $\varepsilon_1$ modulates the feedback from the oscillators to the environment. The damping parameter $\lambda_{\mathrm{env}}$ models the intrinsic decay of the environmental variable in the absence of external input. To introduce a single change point at $t=10$ seconds, the simulation is divided into two phases:
\begin{itemize}
    \item \textbf{Phase 1 (Coupled dynamics):} Active coupling with $\varepsilon_1=0.5$, $\varepsilon_2=0.9$, and $c=24.0$.
    \item \textbf{Phase 2 (Decoupled dynamics):} Coupling is disabled by setting $\varepsilon_1=\varepsilon_2=0$, and the internal parameter is adjusted to $c=27.0$.
\end{itemize}

This creates a structural change in the dynamics of the system, while keeping the observation dimensionality constant. The initial conditions for $x_p(0)$, $y_p(0)$, and $z_p(0)$ are drawn uniformly from $[-1,1]$, and $w(0)=0$. The system is simulated with a time step of $\Delta t=0.01$.

For numerical integration, we employ the \textit{Runge--Kutta method of order 5(4) (RK45)}~\cite{Runge-Kutta} from the \textit{SciPy} library using the \texttt{solve\_ivp} function. To reduce the influence of initial transient dynamics and focus on the steady-state, we discard the first $\tau=100$ time steps $(1s)$, following the practice in the study of environmentally coupled chaotic systems~\cite{coupling}. The resulting dataset captures rich, non-stationary chaotic dynamics, providing a challenging benchmark for evaluating change point detection algorithms.

\subsection{Crowd Simulation}
\label{sec:crowdsim}
In this study, we use a crowd simulation dataset described in \cite{ours} created using Unity\textsuperscript{\textregistered}, a cross-platform game engine developed by Unity Technologies. The dataset includes both normal and abnormal crowd behavior scenarios: a train station and a bi-directional corridor collision scenario. Both simulations feature configurable parameters such as agent count, arrival timing, and spawn rate, allowing for flexible generation of crowd dynamics under controlled settings.

\subsubsection{Overview of Scenarios} The \textit{Train Station Scenario} models typical commuter behavior in a train station environment (Fig.~\ref{fig:train_scenario}). In the \textit{normal} case, agents follow a predictable pattern: entering through the gates, moving toward the train doors, and exiting the station. The \textit{abnormal} case introduces an object in the environment that triggers an emergency response. Once the object appears, the agents immediately deviate from their planned paths and evacuate through the exit, simulating realistic panic-induced behavior.

In \textit{Bi-directional Corridor} scenario, a group of agents moves along a corridor from one end to the other (Fig.~\ref{fig:collision_scenario}). In the \textit{standard} setting, agents follow their intended trajectory without interference. In the \textit{abnormal} setting, a second group is introduced from the opposite end, resulting in intersecting paths that lead to collisions followed by chaotic dispersal. This scenario tests the model’s sensitivity to abrupt, localized disruptions in collective motion.

To introduce behavioral diversity in both scenarios, we define two agent types:
\begin{itemize}
    \item \textbf{Active agents:} Navigate toward defined goals and generate accelerometer data.
    \item \textbf{Passive agents:} Move randomly within the scene and serve as dynamic obstacles, increasing environmental complexity and noise.
\end{itemize}
This mixture increases realism by mimicking both purposeful and background motion within crowds.

\subsubsection{Data Recording and Post-Processing}
For each simulation, we record the three-dimensional position ($x$, $y$, $z$) of the upper left leg of the active agent at each timestep. To reduce noise and simulation artifacts, the data are smoothed using a Savitzky-Golay filter \cite{sav-gol}, which preserves the shape of trajectories while reducing jitter. We then apply a second-order numerical differentiator to estimate acceleration, approximating the data captured by real-world inertial sensors, such as those found in smartphones. These derived features are essential for detecting subtle deviations in motion patterns. 

\section{Experiments}
\label{sec: experiment}

\subsection{Evaluation Metrics}
\label{sec: evaluation_metrics}

To assess the performance of our CPD method, we define evaluation metrics based on the relationship between the actual change point, $t^*$, and the detected change point, $\hat{t}$. Let $N_a$ denote the number of test cases with an actual change point, and $N_n$ denote the number of test cases without a change point. The metrics are defined as follows:

\paragraph{\textbf{Correct Detection (CD)}} A detection is considered correct if the algorithm signals a change at or after the true change, i.e., $\hat{t} \geq t^*$. This corresponds to true positive (TP), which counts the number of times a change is correctly detected. The proportion of correctly detected change points is calculated as: $CD=\text{TP}/N_a$.
\paragraph{\textbf{Missed Detection (MD)}} A change is missed if no detection occurs after the true change, i.e., $\hat{t}=\infty$. This corresponds to a false negative (FN), which counts the number of times an actual change was not detected. The proportion of missed changes is $MD=\text{FN}/N_a$.
\paragraph{\textbf{False Alarm (FA)}} A false alarm occurs when the algorithm signals a change before the true change $\hat{t}<t^*$ or in cases where there is no actual change. This corresponds to a false positive (FP), which counts the number of incorrect detections. The false alarm rate is $FA= \text{FP}/(N_a + N_n)$.
\paragraph{\textbf{Detection Delay (DD)}} The latency between the true change point and the detection by the algorithm. Let $t^*$ be the timestamp of the actual change, and let $\hat{t}_{i} > t^*$ denote the detection time in the $i$-th simulation. 

To analyze detection delay, we use the cumulative distribution of delays across simulations where a change point was successfully detected by our method. The sampling interval of the underlying time series is denoted by $\Delta t$. For each simulation $i=1,2,\dots,N_s$, the detection delay is calculated as:
\begin{equation}
    d_i = (\hat{t}_{i} - t^*) \cdot \Delta t
\end{equation}

The full set of delays across all simulations is:
\begin{equation}
    D=\{d_i\}^{N_s}_{i=1}.
\end{equation}

Let $f(d)$ denote the frequency (count) of each delay $d \in D$, and let the total number of valid detections, true positives, be:
\begin{equation}
    N = \sum_d f(d).
\end{equation}
We define a resolution grid:
\begin{equation}
    \mathcal{R} = 
        \left\{ r_j = d_{\min} + (j-1)\Delta t\right\}_{j=0}^{N_d}
\end{equation}
where 
\begin{equation}
    N_d = \lfloor (d_{\max} - d_{\min})/\Delta t\rfloor
\end{equation}
and $d_{max}$ and $d_{min}$ and the largest and smallest observed delay values, respectively. Using this grid, the cumulative distribution function (CDF) of the detection delays is defined as
\begin{equation}
    F_D(r_j) = \mathbb{P}(D \leq r_j) = \sum_{d \leq r_j} \frac{f(d)}{N}
\end{equation}
For a given confidence level $\alpha \in [0, 1]$, the detection delay at that confidence is defined as:
\begin{equation}
    d_{\alpha} = \min \left\{ r_j \in \mathcal{R} \,\middle|\, F_D(r_j) \geq \alpha \right\}
\end{equation}

\subsection{Anomaly Scoring Models at Entity Level}
\label{sec:mtsad}

Multivariate time series anomaly detection has received significant research attention, leading to models capable of capturing complex temporal dependencies and variable interactions at the entity level. This study includes several deep learning MTS-AD models to compute reconstruction-based anomaly scores for each entity.

\begin{figure*}
\centering
\includegraphics[keepaspectratio, width=\textwidth]{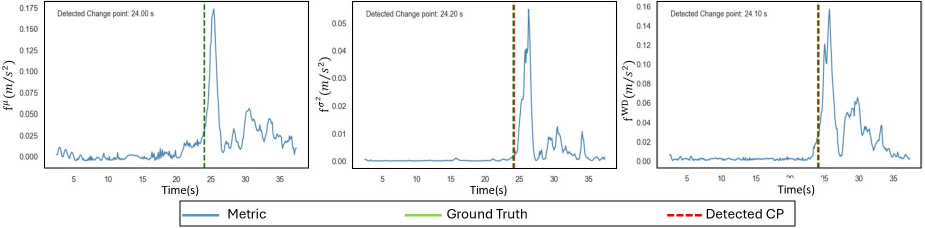}
\caption{\textbf{Change point detection using aggregated IDfN scores across entities.} \textbf{From left to right:} Aggregated metrics—Mean ($f^{\mu}$), Variance ($f^{\sigma^2}$), and Kernel Density-based Wasserstein Distance ($f^{\mathrm{WD}}$)—which highlight a systemic change. The ground truth change point is shown in green. The dashed red line denotes the change point detected by our proposed method, based on each aggregation strategy. All three aggregation strategies detect the change point in close proximity to the ground truth.}
\label{fig:metrics}
\end{figure*}

\begin{itemize}
    \item \textbf{Proposed Model} We use a Simple Autoencoder (SAE) composed of fully connected, dense layers with ReLU activations. The architecture is designed to compress and reconstruct multivariate time series data. This model serves as a basic deep learning approach in our study.
    \item \textbf{OmniAnomaly} \cite{Omni} Combines a variational autoencoder with stochastic recurrent neural networks to model temporal dependencies and uncertainty in multivariate time series. It learns a probabilistic latent representation of normal patterns and assigns anomaly scores based on reconstruction likelihood. By optimizing both the reconstruction error and the distributional divergence, it captures complex structured deviations in sequential data.
    \item \textbf{USAD} \cite{usad}: Utilizes a dual autoencoder structure with a shared encoder and two decoders to learn patterns in multivariate time series without labeled anomalies. Train in two phases: a conventional autoencoder phase and an adversarial phase that amplifies reconstruction errors for anomalous data.
    \item \textbf{TranAD} \cite{tranAD}: A Transformer-based anomaly detection model designed to address the limitations of recurrent neural networks by capturing both short- and long-range dependencies in time series. Using self-attention, it assigns dynamic weights to each time step, highlighting crucial parts of the sequence to detect subtle anomalies.
\end{itemize}

These models\footnote{We use publicly available code sources for all of the baselines implemented in https://github.com/imperial-qore/TranAD} provide diverse strategies for anomaly scoring, which we use as input to our system-wide change point detection framework.

\subsection{Training Setup}

Each model is trained with Adam optimizer \cite{adam}, except for TranAD, which follows its original implementation with AdamW to improve weight regularization and adaptive learning. For SAE, we minimize MSE loss with an initial learning rate of $1 \times 10^{-4}$. To facilitate efficient convergence, we dynamically adjust the learning rate using the ReduceLROnPlateau callback, reducing it by a factor of $0.5$ if the validation loss stagnates for five consecutive epochs. The learning rate is constrained by a lower bound of $1 \times 10^{-6}$. Training is performed with a batch size of $512$ for a maximum of $20$ epochs, incorporating early stopping if the validation loss does not improve after five epochs. Under these conditions, SAE typically converges within approximately $10$ epochs. To mitigate overfitting, we apply a dropout rate of $0.1$ after each dense layer. Additionally, we fine-tune key hyperparameters, such as the number of hidden units and the dropout rate, using grid search to optimize performance.  

For other models, hyperparameter tuning determines an optimal initial learning rate of $1 \times 10^{-3}$. Same as SAE, we find that training these models for $10$ epochs is sufficient to achieve competitive performance. All models are trained using early stopping and learning rate scheduling to balance training efficiency and generalization. These configurations ensure stable convergence while preventing overfitting.

\begin{table*}[t]
\centering
\caption{Change Point Detection Results (CUSUM - Mean) across Models and Datasets.}
\renewcommand{\arraystretch}{1.1}
\resizebox{\textwidth}{!}{
\begin{tabular}{|c|cccc|cccc|cccc|cccc|cccc|}
\hline
\multirow{3}{*}{\textbf{Model}} 
& \multicolumn{4}{c|}{\multirow{2}{*}{\textbf{Crowd-Station}}} 
& \multicolumn{4}{c|}{\multirow{2}{*}{\textbf{Crowd-Collision}}} 
& \multicolumn{4}{c|}{\multirow{2}{*}{\textbf{Coupled Chen}}} 
& \multicolumn{8}{c|}{\textbf{AR}} \\
\cline{14-21}
& & & & & & & & & & & & 
& \multicolumn{4}{c|}{\textbf{Change in Mean}} 
& \multicolumn{4}{c|}{\textbf{Change in Variance}} \\
\cline{2-21}
& \textbf{CD} & \textbf{DD(s)} & \textbf{FA} & \textbf{MD} 
& \textbf{CD} & \textbf{DD(s)} & \textbf{FA} & \textbf{MD} 
& \textbf{CD} & \textbf{DD(s)} & \textbf{FA} & \textbf{MD} 
& \textbf{CD} & \textbf{DD(s)} & \textbf{FA} & \textbf{MD} 
& \textbf{CD} & \textbf{DD(s)} & \textbf{FA} & \textbf{MD} \\
\hline
TranAD      
& 0.99 & 0.4 & 0.01 & 0.0 
& \textbf{1.0} & 0.6 & \textbf{0.0} & 0.0  
& 0.93 & 1.2 & 0.02 & 0.05   
& 0.96 & 2.0 & 0.04 & 0.0  
& 0.98 & 4.0 & 0.02 & 0.0  \\
USAD        
& \textbf{1.0} & 0.4 & \textbf{0.0} & 0.0 
& \textbf{1.0} & \textbf{0.5} & \textbf{0.0} & 0.0  
& 0.90 & 1.2 & 0.05 & 0.05   
& 0.96 & 2.0 & 0.04 & 0.0  
& 0.97 & 4.0 & 0.03 & 0.0  \\
OmniAnomaly 
& 0.98 & 0.5 & 0.02 & 0.0 
& 0.99 & \textbf{0.5} & 0.01 & 0.0  
& 0.90 & 1.2 & 0.05 & 0.05
& 0.96 & 1.0 & 0.04 & 0.0  
& 0.98 & 5.0 & 0.02 & 0.0    \\
Ours        
& \textbf{1.0} & \textbf{0.2} & \textbf{0.0} & 0.0 
& \textbf{1.0} & \textbf{0.5} & \textbf{0.0} & 0.0  
& \textbf{0.98} & \textbf{0.2} & \textbf{0.02} & \textbf{0.0}   
& \textbf{1.0} & \textbf{1.0} & \textbf{0.0} & 0.0  
& \textbf{1.0} & \textbf{3.0} & \textbf{0.0} & 0.0  \\
\hline
\end{tabular}
}
\label{tab:cpd_mean}
\end{table*}

\begin{table*}[t]
\centering
\caption{Change Point Detection Results (CUSUM - Variance) across Models and Datasets.}
\renewcommand{\arraystretch}{1.1}
\resizebox{\textwidth}{!}{
\begin{tabular}{|c|cccc|cccc|cccc|cccc|cccc|}
\hline
\multirow{3}{*}{\textbf{Model}}
& \multicolumn{4}{c|}{\multirow{2}{*}{\textbf{Crowd-Station}}} 
& \multicolumn{4}{c|}{\multirow{2}{*}{\textbf{Crowd-Collision}}} 
& \multicolumn{4}{c|}{\multirow{2}{*}{\textbf{Coupled Chen}}} 
& \multicolumn{8}{c|}{\textbf{AR}} \\
\cline{14-21}
& & & & & & & & & & & & 
& \multicolumn{4}{c|}{\textbf{Change in Mean}} 
& \multicolumn{4}{c|}{\textbf{Change in Variance}} \\
\cline{2-21}
& \textbf{CD} & \textbf{DD(s)} & \textbf{FA} & \textbf{MD} 
& \textbf{CD} & \textbf{DD(s)} & \textbf{FA} & \textbf{MD} 
& \textbf{CD} & \textbf{DD(s)} & \textbf{FA} & \textbf{MD} 
& \textbf{CD} & \textbf{DD(s)} & \textbf{FA} & \textbf{MD} 
& \textbf{CD} & \textbf{DD(s)} & \textbf{FA} & \textbf{MD} \\
\hline
TranAD      
& 0.99 & 0.9 & 0.01 & 0.0 
& \textbf{0.95} & 1.2 & \textbf{0.04} & 0.01  
& 0.95 & 1.2 & 0.04 & 0.01   
& 0.97 & 4.0 & 0.03 & 0.0  
& 0.98 & 5.0 & 0.02 & 0.0  \\
USAD        
& 0.98 & 0.9 & 0.02 & 0.0 
& 0.94 & 1.0 & 0.06 & \textbf{0.0}  
& 0.93 & 1.2 & 0.06 & 0.01   
& 0.96 & 4.0 & 0.04 & 0.0  
& 0.98 & \textbf{4.0} & 0.02 & 0.0  \\
OmniAnomaly 
& 0.98 & 0.9 & 0.02 & 0.0 
& 0.94 & 1.1 & 0.06 & \textbf{0.0} 
& 0.94 & 1.0 & 0.06 & \textbf{0.0}   
& 0.94 & 3.0 & 0.06 & 0.0      
& 0.96 & \textbf{4.0} & 0.04 & 0.0  \\
Ours        
& \textbf{1.0} & \textbf{0.7} & \textbf{0.0} & 0.0 
& 0.94 & \textbf{0.6} & 0.06 & \textbf{0.0}  
& \textbf{1.0} & \textbf{0.3} & \textbf{0.0} & \textbf{0.0}   
& \textbf{1.0} & \textbf{2.0} & \textbf{0.0} & 0.0  
& \textbf{1.0} & \textbf{4.0} & \textbf{0.0} & 0.0  \\
\hline
\end{tabular}
}
\label{tab:cpd_variance}
\end{table*}

\begin{table*}[t]
\centering
\caption{Change Point Detection Results (CUSUM - Wass+KDE) across Models and Datasets.}
\renewcommand{\arraystretch}{1.1}
\resizebox{\textwidth}{!}{%
\begin{tabular}{|c|cccc|cccc|cccc|cccc|cccc|}
\hline
\multirow{3}{*}{\textbf{Model}} 
& \multicolumn{4}{c|}{\multirow{2}{*}{\textbf{Crowd-Station}}} 
& \multicolumn{4}{c|}{\multirow{2}{*}{\textbf{Crowd-Collision}}} 
& \multicolumn{4}{c|}{\multirow{2}{*}{\textbf{Coupled Chen}}} 
& \multicolumn{8}{c|}{\textbf{AR}} \\
\cline{14-21}
& & & & & & & & & & & & 
& \multicolumn{4}{c|}{\textbf{Change in Mean}} 
& \multicolumn{4}{c|}{\textbf{Change in Variance}} \\
\cline{2-21}
& \textbf{CD} & \textbf{DD(s)} & \textbf{FA} & \textbf{MD} 
& \textbf{CD} & \textbf{DD(s)} & \textbf{FA} & \textbf{MD} 
& \textbf{CD} & \textbf{DD(s)} & \textbf{FA} & \textbf{MD} 
& \textbf{CD} & \textbf{DD(s)} & \textbf{FA} & \textbf{MD} 
& \textbf{CD} & \textbf{DD(s)} & \textbf{FA} & \textbf{MD} \\
\hline
TranAD      
& 0.94 & 0.6 & 0.04 & 0.02 
& \textbf{1.0} & 1.2 & \textbf{0.0} & 0.0  
& 0.88 & 1.0 & 0.06 & 0.06   
& 0.99 & 3.0 & \textbf{0.0} & 0.01  
& 0.95 & 8.0 & 0.05 & 0.0  \\
USAD        
& 0.94 & \textbf{0.5} & 0.05 & 0.01 
& 0.99 & 1.0 & 0.01 & 0.0  
& 0.82 & 1.0 & 0.09 & 0.09   
& 0.96 & 3.0 & 0.00 & 0.04  
& 0.94 & 8.0 & 0.06 & 0.0  \\
OmniAnomaly 
& 0.95 & \textbf{0.5} & 0.05 & \textbf{0.0} 
& \textbf{1.0} & 1.1 & \textbf{0.0} & 0.0  
& 0.82 & 1.1 & 0.07 & 0.05
& 0.95 & 2.0 & 0.04 & 0.01  
& 0.95 & 9.0 & 0.05 & 0.0    \\
Ours        
& \textbf{1.0} & 0.6 & \textbf{0.0} & \textbf{0.0} 
& 0.98 & \textbf{0.5} & 0.02 & 0.0  
& \textbf{1.0} & \textbf{0.2} & \textbf{0.0} & \textbf{0.0}   
& \textbf{1.0} & \textbf{1.0} & \textbf{0.0} & \textbf{0.0}  
& \textbf{1.0} & \textbf{3.0} & \textbf{0.0} & 0.0  \\
\hline
\end{tabular}
}
\label{tab:cpd_kde}
\end{table*}

\section{Results}
\label{sec: result}

\subsection{CPD Performance Results}

In this section, we present the experimental evaluation of our CPD framework.
The framework applies CUSUM-based detectors to three different statistics: mean, variance, and Wasserstein distance with KDE (Wass+KDE). The goal is to systematically compare the effectiveness of each statistic across diverse datasets and change scenarios. Detailed quantitative results are presented in Tables~\ref{tab:cpd_mean}–\ref{tab:cpd_kde}. To complement this quantitative analysis, Figure~\ref{fig:metrics} illustrates the aggregated detection scores, \textit{SWAS}, over time in the Crowd-Train dataset.

In the following, we discuss the results for each scoring statistic separately, highlighting key trends and insights.

\subsubsection{Results: Mean-based Detection} 

Table~\ref{tab:cpd_mean} summarizes the performance of CPD detection when applying CUSUM on the mean of IDfN scores across different datasets and models.
Our proposed method consistently achieves the highest correct detection, with perfect scores of 1.00 on the Crowd-Station, Crowd-Collision, and AR datasets, and near-perfect performance on the Coupled Chen dataset (CD = 0.98). This indicates a robust sensitivity to changes that affect the mean behavior of the underlying data.
In terms of detection delay, our approach also demonstrates superior promptness. For example, the delay on the Crowd-Station and Coupled Chen datasets is significantly lower (0.2 seconds) compared to baselines such as TranAD and USAD, which have delays of up to 1.2 seconds on the Coupled Chen dataset. Faster detection is crucial for timely interventions in practical applications.

False alarm rates are minimal or zero in all data sets, highlighting the precision of the method in avoiding spurious change detections. Similarly, missed detections are consistently zero or near zero, confirming reliability.
Baseline methods, such as TranAD, USAD, and OmniAnomaly, perform well overall but exhibit slightly higher delays and false alarms, particularly on the more complex Coupled Chen dataset.
Overall, these results demonstrate that leveraging the mean of reconstruction errors with our adaptive CUSUM framework yields a highly accurate and responsive change point detection system across diverse, multi-entity time series datasets.

\subsubsection{Results: Variance-based Detection} 
Table~\ref{tab:cpd_variance} summarizes the performance of CPD using variance as a global aggregation method. Across most of the datasets, our method achieves the highest correct detection rates, reaching 1.00 on Crowd-Station, Coupled Chen, and both AR datasets. It also consistently shows the lowest detection delay, with delays as low as 0.3 seconds on Coupled Chen and 0.6 seconds on Crowd-Collision. In contrast, TranAD, USAD, and OmniAnomaly typically have higher detection delays, often ranging from 0.9 to 1.2 seconds on these datasets.
Competing methods have small but non-zero FA values (e.g., between 0.01 and 0.06), while our method keeps FA at 0.0 on nearly all datasets. This demonstrates greater robustness to noise in the variance signal. Finally, in AR datasets, our method outperforms the baselines by achieving perfect correct detection and zero false alarms, along with shorter detection delays (e.g., DD = 2.0 seconds for the change in mean scenario, compared to 4.0 to 5.0 seconds for baselines).

Overall, these results show that the variance-based aggregation method, applied to the Coupled Chen dataset, enhances detection accuracy and reduces missed detections compared to the mean-based approach, while maintaining a comparable detection delay.

\begin{figure*}
\centering
\includegraphics[keepaspectratio, width=\textwidth]{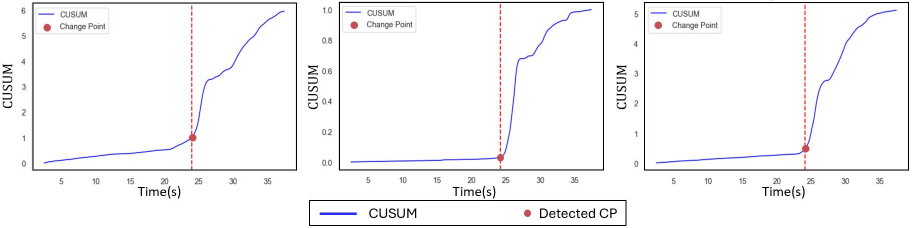}
\caption{\textbf{CUSUM-based change point detection over aggregated IDfN metrics.} \textbf{From left to right:} CUSUM computed over the Mean ($f^{\mu}$), Variance ($f^{\sigma^2}$), and Kernel Density-based Wasserstein Distance ($f^{\mathrm{WD}}$) of aggregated scores. The CUSUM statistic at time $t$ is defined recursively as shown in Equation~\ref{eq:cusum}. The blue curve represents the CUSUM statistic. Detected change points based on the proposed method are marked with a red dashed line and a dot. All three metrics yield timely detection of the underlying change.}
\label{fig:cusum}
\end{figure*}

\begin{table*}[t]
\centering
\caption{Change Point Detection Results (Ours Only) across Detection Methods and Datasets.}
\renewcommand{\arraystretch}{1.1}
\resizebox{\textwidth}{!}{
\begin{tabular}{|c|cccc|cccc|cccc|cccc|cccc|}
\hline
\multirow{3}{*}{\textbf{Method}} &
    \multicolumn{4}{c|}{\multirow{2}{*}{\textbf{Crowd-Station}}} & 
    \multicolumn{4}{c|}{\multirow{2}{*}{\textbf{Crowd-Collision}}} &
    \multicolumn{4}{c|}{\multirow{2}{*}{\textbf{Coupled Chen}}} &
    \multicolumn{8}{c|}{\textbf{AR}} \\ \cline{14-21}
  & \multicolumn{4}{c|}{} &
    \multicolumn{4}{c|}{} &
    \multicolumn{4}{c|}{} &
    \multicolumn{4}{c|}{\textbf{Change in Mean}} &
    \multicolumn{4}{c|}{\textbf{Change in Variance}} \\ \cline{2-21}
& \textbf{CD} & \textbf{DD(s)} & \textbf{FA} & \textbf{MD} 
& \textbf{CD} & \textbf{DD(s)} & \textbf{FA} & \textbf{MD} 
& \textbf{CD} & \textbf{DD(s)} & \textbf{FA} & \textbf{MD} 
& \textbf{CD} & \textbf{DD(s)} & \textbf{FA} & \textbf{MD} 
& \textbf{CD} & \textbf{DD(s)} & \textbf{FA} & \textbf{MD} \\
\hline
CUSUM + Variance 
& 1.00 & 0.7 & 0.00 & 0.00 
& 0.94 & 0.6 & 0.06 & 0.00  
& 1.00 & 0.3 & 0.00 & 0.00   
& 1.00 & 2.0 & 0.00 & 0.00  
& 1.00 & 4.0 & 0.00 & 0.00  \\
CUSUM + Mean    
& 1.00 & 0.2 & 0.00 & 0.00 
& 1.00 & 0.5 & 0.00 & 0.00  
& 0.98 & 0.2 & 0.02 & 0.00   
& 1.00 & 1.0 & 0.00 & 0.00  
& 1.00 & 3.0 & 0.00 & 0.00  \\
CUSUM + Wass+KDE 
& 1.00 & 0.6 & 0.00 & 0.00 
& 0.98 & 0.5 & 0.02 & 0.00  
& 1.00 & 0.2 & 0.00 & 0.00   
& 1.00 & 1.0 & 0.00 & 0.00  
& 1.00 & 3.0 & 0.00 & 0.00  \\
\hline
\end{tabular}
}
\label{tab:cpd_ours}
\end{table*}

\subsubsection{Results: Wasserstein+KDE-based Detection} 

Table~\ref{tab:cpd_kde} reports the performance of the proposed method using the CUSUM detector applied to the Wasserstein distance combined with KDE across all datasets and models.

Our method achieves perfect or near-perfect correct detection rates across most datasets, reaching 1.00 on Crowd-Station, Coupled Chen, and AR datasets for both changes in mean and variance. This confirms the robustness of using distributional distance measures for detecting subtle and complex changes in time series.
Detection delays are competitive, especially on the Coupled Chen dataset (0.2 seconds) and AR dataset (1.0 to 3.0 seconds), significantly outperforming TranAD, USAD, and OmniAnomaly, which experience delays up to 8.0 seconds on AR variance changes. The low delay highlights the responsiveness of the Wass+KDE approach.
False alarm rates remain very low or zero for our method, showcasing precise identification of true change points. Missed detection rates are also minimal or zero, emphasizing reliability even for challenging cases such as the Coupled Chen dataset, where other methods show elevated missed detections.
While baseline models demonstrate solid results, they occasionally suffer from higher missed detection rates and false alarms, particularly on datasets with complex change dynamics.
Overall, the Wass + KDE statistic, when combined with our adaptive detection framework, proves to be a powerful tool for identifying change points with high accuracy, low latency, and strong robustness to different types of distributional changes.

Figure~\ref{fig:cusum} shows the CUSUM statistics over time for aggregated detection scores (mean, variance, and Wass+KDE). These visualizations illustrate how each detector accumulates evidence, explaining differences in detection delay and sensitivity seen in the tables. 

\subsection{Analysis and Discussion}

The experimental results across the three statistical approaches—mean, variance, and Wass+KDE—highlight the effectiveness and adaptability of our CPD framework. Across all datasets and change types, our method consistently achieves high correct detection rates, often outperforming or closely matching state-of-the-art baselines, such as TranAD, USAD, and OmniAnomaly, while maintaining a lower training time per epoch \cite{ours}. This consistency underscores the robustness of combining reconstruction errors from our proposed model with CUSUM-based detectors. For clarity, Table~\ref{tab:cpd_ours} presents the results of our proposed model across all datasets, highlighting its strong performance independent of the chosen statistic. Among the three statistics, the mean-based detector generally demonstrates the strongest and most stable performance. It achieves near-perfect correct detection rates with very low detection delays and minimal false alarms, even on the challenging dataset, Coupled Chen. The low detection latency is particularly advantageous for real-time applications where a timely response is critical. The variance-based detector performs competitively, particularly when changes involve variance, though it can sometimes exhibit slightly higher false alarm rates.

One of the most notable findings is that our method consistently yields lower detection delays compared to baseline models, indicating a quicker response across all tested statistics. This responsiveness is crucial in practical settings where early detection can enable faster decision-making and intervention. Additionally, our framework maintains low false alarm and miss detection rates, underscoring its reliability and precision across diverse data conditions. 

From a practical perspective, these results suggest that researchers can improve change detection by selecting the most suitable statistic based on domain knowledge or by combining multiple statistics for greater robustness. While the Wass+KDE method introduces higher computational complexity, it can remain beneficial when detecting subtle or higher-order distributional changes is important.

In summary, our comprehensive evaluation shows that while mean- and variance-based detectors provide strong baseline performance, integrating distribution-sensitive metrics, such as Wass+KDE, further enhances detection accuracy and timeliness for some datasets. This confirms the flexibility, adaptability, and robustness of our CPD framework, making it potentially well-suited for a wide range of real-world change detection applications.

\begin{table*}[t]
\centering
\caption{Anomaly Detection Results across Different Detection Scores and Datasets.}
\renewcommand{\arraystretch}{1.1}
\resizebox{\textwidth}{!}{
\begin{tabular}{|cc|cc|cc|cc|cc|cc|}
\hline
\multicolumn{2}{|c|}{\multirow{3}{*}{\textbf{Method}}} &
  \multicolumn{2}{c|}{\multirow{2}{*}{\textbf{Crowd-Station}}} &
  \multicolumn{2}{c|}{\multirow{2}{*}{\textbf{Crowd-Collision}}} &
  \multicolumn{2}{c|}{\multirow{2}{*}{\textbf{Coupled Chen}}} &
  \multicolumn{4}{c|}{\textbf{AR}} \\ \cline{9-12} 
\multicolumn{2}{|c|}{} &
  \multicolumn{2}{c|}{} &
  \multicolumn{2}{c|}{} &
  \multicolumn{2}{c|}{} &
  \multicolumn{2}{c|}{\textbf{Change in Mean}} &
  \multicolumn{2}{c|}{\textbf{Change in Variance}} \\ \cline{3-12} 
\multicolumn{2}{|c|}{} &
  \multicolumn{1}{c|}{\textbf{AUROC}} &
  \textbf{F1-Score} &
  \multicolumn{1}{c|}{\textbf{AUROC}} &
  \textbf{F1-Score} &
  \multicolumn{1}{c|}{\textbf{AUROC}} &
  \textbf{F1-Score} &
  \multicolumn{1}{c|}{\textbf{AUROC}} &
  \multicolumn{1}{c|}{\textbf{F1-Score}} &
  \multicolumn{1}{c|}{\textbf{AUROC}} &
  \textbf{F1-Score} \\ \hline
\multicolumn{1}{|c|}{\multirow{4}{*}{\textbf{Mean}}} &
  TranAD &
  \multicolumn{1}{c|}{0.989} &
  0.944 &
  \multicolumn{1}{c|}{0.993} &
  0.960 &
  \multicolumn{1}{c|}{0.832} &
  0.785 &
  \multicolumn{1}{c|}{0.994} &
  0.993 &
  \multicolumn{1}{c|}{0.991} &
  0.978 \\ 
\multicolumn{1}{|c|}{} &
  USAD &
  \multicolumn{1}{c|}{0.990} &
  0.948 &
  \multicolumn{1}{c|}{0.991} &
  0.958 &
  \multicolumn{1}{c|}{0.839} &
  0.769 &
  \multicolumn{1}{c|}{0.994} &
  0.992 &
  \multicolumn{1}{c|}{0.991} &
  0.977 \\ 
\multicolumn{1}{|c|}{} &
  OmniAnomaly &
  \multicolumn{1}{c|}{0.990} &
  0.944 &
  \multicolumn{1}{c|}{0.992} &
  0.958 &
  \multicolumn{1}{c|}{0.837} &
  0.769 &
  \multicolumn{1}{c|}{\textbf{0.999}} &
  0.996 &
  \multicolumn{1}{c|}{0.990} &
  \textbf{0.980} \\ 
\multicolumn{1}{|c|}{} &
  Ours &
  \multicolumn{1}{c|}{\textbf{0.993}} &
  \textbf{0.954} &
  \multicolumn{1}{c|}{\textbf{0.999}} &
  \textbf{0.989} &
  \multicolumn{1}{c|}{\textbf{0.915}} &
  \textbf{0.879} &
  \multicolumn{1}{c|}{0.998} &
  \textbf{0.997} &
  \multicolumn{1}{c|}{\textbf{0.992}} &
  \textbf{0.980} \\ \hline

\multicolumn{1}{|c|}{\multirow{4}{*}{\textbf{Variance}}} &
  TranAD &
  \multicolumn{1}{c|}{0.972} &
  0.889 &
  \multicolumn{1}{c|}{0.993} &
  0.970 &
  \multicolumn{1}{c|}{0.855} &
  0.762 &
  \multicolumn{1}{c|}{0.991} &
  0.970 &
  \multicolumn{1}{c|}{0.989} &
  0.966 \\ 
\multicolumn{1}{|c|}{} &
  USAD &
  \multicolumn{1}{c|}{0.972} &
  0.889 &
  \multicolumn{1}{c|}{0.993} &
  0.971 &
  \multicolumn{1}{c|}{0.860} &
  0.790 &
  \multicolumn{1}{c|}{0.990} &
  0.968 &
  \multicolumn{1}{c|}{0.990} &
  0.967 \\ 
\multicolumn{1}{|c|}{} &
  OmniAnomaly &
  \multicolumn{1}{c|}{0.970} &
  0.888 &
  \multicolumn{1}{c|}{0.992} &
  0.970 &
  \multicolumn{1}{c|}{0.858} &
  0.788 &
  \multicolumn{1}{c|}{\textbf{0.997}} &
  0.977 &
  \multicolumn{1}{c|}{0.990} &
  \textbf{0.969} \\ 
\multicolumn{1}{|c|}{} &
  Ours &
  \multicolumn{1}{c|}{\textbf{0.983}} &
  \textbf{0.919} &
  \multicolumn{1}{c|}{\textbf{0.998}} &
  \textbf{0.983} &
  \multicolumn{1}{c|}{\textbf{0.901}} &
  \textbf{0.895} &
  \multicolumn{1}{c|}{0.996} &
  \textbf{0.983} &
  \multicolumn{1}{c|}{\textbf{0.992}} &
  0.966 \\ \hline

\multicolumn{1}{|c|}{\multirow{4}{*}{\textbf{Wass+KDE}}} &
  TranAD &
  \multicolumn{1}{c|}{0.982} &
  0.922 &
  \multicolumn{1}{c|}{0.945} &
  0.910 &
  \multicolumn{1}{c|}{0.758} &
  0.720 &
  \multicolumn{1}{c|}{0.985} &
  0.976 &
  \multicolumn{1}{c|}{0.939} &
  0.911 \\ 
\multicolumn{1}{|c|}{} &
  USAD &
  \multicolumn{1}{c|}{0.983} &
  0.927 &
  \multicolumn{1}{c|}{0.944} &
  0.917 &
  \multicolumn{1}{c|}{0.774} &
  0.731 &
  \multicolumn{1}{c|}{0.985} &
  0.976 &
  \multicolumn{1}{c|}{0.942} &
  0.914 \\ 
\multicolumn{1}{|c|}{} &
  OmniAnomaly &
  \multicolumn{1}{c|}{0.982} &
  0.921 &
  \multicolumn{1}{c|}{0.945} &
  0.911 &
  \multicolumn{1}{c|}{0.772} &
  0.730 &
  \multicolumn{1}{c|}{0.991} &
  0.977 &
  \multicolumn{1}{c|}{0.940} &
  0.911 \\ 
\multicolumn{1}{|c|}{} &
  Ours &
  \multicolumn{1}{c|}{\textbf{0.990}} &
  \textbf{0.942} &
  \multicolumn{1}{c|}{\textbf{0.999}} &
  \textbf{0.987} &
  \multicolumn{1}{c|}{\textbf{0.850}} &
  \textbf{0.840} &
  \multicolumn{1}{c|}{\textbf{0.997}} &
  \textbf{0.997} &
  \multicolumn{1}{c|}{\textbf{0.991}} &
  \textbf{0.972} \\ \hline
\end{tabular}
}
\label{tab:auroc_f1_results}
\end{table*}

\begin{figure}[!t]
    \centering
    \includegraphics[width=0.5\textwidth]{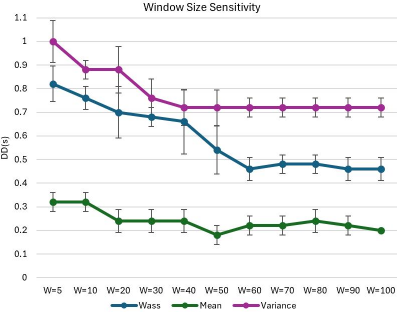}
    \caption{The average detection delay (in seconds) as a function of sliding window size $W$ for three statistics used within our method: Wass+KDE (blue), mean (green), and variance (pink). The plot displays the average detection delay across five repeated runs on the crowd simulation dataset, with error bars representing the standard deviation.}
    \label{fig:delay_vs_window}
\end{figure}

\subsection{Window Size Sensitivity}
The window size parameter $W$ plays a critical role in CPD for time series, as it determines the amount of historical data used to compute detection statistics at each time step~\cite{window-size}. To evaluate the impact of $W$ on detection performance, we tested a range of window sizes $W = \{5, 10, 20, \ldots, 100\}$ on the crowd-train dataset. For each setting, we computed the detection delay across five repeated runs. Figure~\ref{fig:delay_vs_window} illustrates how detection delay varies with window size for three statistics used in our method: mean, variance, and Wass+KDE. The mean statistic consistently achieves the lowest detection delay. It remains relatively stable across window sizes, reflecting its robustness to noise and rapid responsiveness.

In contrast, the Wass+KDE statistic shows higher detection delay at smaller window sizes ($W \leq 20$), but its performance improves significantly and becomes more stable when $W \geq 40$. The variance statistic exhibits the highest detection delays and larger fluctuations, likely due to its sensitivity to noise and the instability of variance estimates from small samples.

The observed sensitivity to window size can be attributed to the reconstruction errors $e_{t,p}$ underpinning all statistics. When the window size $W$ is small, the model has a limited historical context to anchor its predictions, leading to higher variability in reconstruction errors~\cite{lastPoint}. Specifically, the model must infer what constitutes "normal" behavior from limited data, resulting in noisier and less reliable estimates. On the other hand, increasing the window size helps the algorithm produce more stable, robust estimates, but also increases computational time~\cite{CPD-time}. Thus, selecting an appropriate window size requires balancing the need for sufficient data to produce stable reconstruction errors to achieve timely detection.

Overall, these results highlight a trade-off between detection speed and sensitivity to window size. The mean statistic provides fast and stable detection with relatively low sensitivity to $W$, whereas the Wass+KDE and variance statistics approach benefits more from larger windows. Based on this analysis, we set the window size to $W=50$ for the crowd simulation dataset, as it offers a good balance between low detection delay and robustness.

\subsection{Anomaly Detection Results}

While the primary focus of this study is on CPD, we also report performance on anomaly detection to provide a comprehensive evaluation that facilitates comparison with prior work \cite{ours}. To this end, we follow the evaluation protocol outlined in \cite{ours}, using the \textit{F1-Score} and the Area Under the Receiver Operating Characteristic Curve (\textit{AUROC}), which effectively assesses classification accuracy and the trade-off between true and false positives \cite{tranAD}. For ground truth labeling, we assign all data points in the non-anomalous dataset the label $0$. In the anomalous dataset, data points before the change point $t^*$ are labeled $0$, and those after are labeled $1$. We perform this evaluation on the scores $f^{\alpha}(t)$, with $\{\alpha \in {\mu, \sigma^2, \text{WD}}\}$.

Table~\ref{tab:auroc_f1_results} summarizes the anomaly detection performance across datasets. Overall, the proposed framework with SAE consistently achieves the highest scores, outperforming TranAD, USAD, and OmniAnomaly under different scoring schemes. In particular, notable gains are observed in the Coupled Chen dataset, where our method improves the AUROC by up to $8\%$ and the F1-Score by more than $10\%$, demonstrating its robustness in more complex multi-entity, multivariate settings. Across AR datasets, all methods report strong performance, yet our approach using SAE maintains a slight but consistent advantage, especially in the variance and Wass+KDE settings. We note, however, that OmniAnomaly achieves the highest AUROC in the AR change-in-mean case, reporting 0.999 for mean metrics and 0.997 for variance metrics, compared to 0.998 and 0.996 for our method. Similarly, in the AR change-in-variance case, OmniAnomaly attains equal or higher F1-scores (0.980 for mean and 0.969 for variance) relative to ours (0.980 and 0.966). These cases highlight the competitiveness of approaches in specific scenarios.

It is important to note that anomaly detection is not the primary focus of this study. Nevertheless, these findings confirm that the proposed framework retains competitive anomaly detection capabilities while being principally designed for efficient and accurate change point detection.

\section{Conclusion}
\label{sec: conclusion}

In this paper, we propose a flexible change point detection framework for multi-entity multivariate time series data that combines reconstruction errors from MTS-AD models with CUSUM-based detectors applied to multiple statistics: mean, variance, and Wasserstein distance with KDE. Extensive experiments demonstrate that the mean-based detector often achieves the strongest and most stable detection performance, with low detection delays and minimal false alarms. The variance-based detector remains effective when changes manifest primarily as a change in variance, while the Wass+KDE approach adds value in capturing more subtle distributional changes, albeit at a higher computational cost. Across all datasets and change scenarios, our method consistently outperforms or closely matches state-of-the-art baselines, while maintaining robustness and low latency.

Overall, our results highlight the practical importance of choosing detection statistics that align with domain characteristics and the types of changes expected. The proposed CPD framework provides a useful, adaptable, and accurate solution well-suited to real-world applications, where timely and reliable identification of structural changes is essential. 

Future work could explore voting-based strategies to combine outputs from different statistics, such as mean, variance, and Wasserstein distance with KDE, which may enhance robustness, reduce false alarms, and leverage their complementary strengths for more reliable CPD. Additionally, incorporating additional modalities, such as Wi-Fi, can provide contextual information and enrich the data, further improving detection accuracy, particularly in situations where primary sensors, like crowdsourced accelerometers, are sparse.

\section*{Acknowledgments}

This material is based upon work supported by the U.S. Department of Homeland Security under Grant Awards 22STESE00001-01-00, 22STESE00001-02-00, and 22STESE00001-03-02. The views and conclusions contained in this document are those of the authors and should not be interpreted as necessarily representing the official policies, either expressed or implied, of the U.S. Department of Homeland Security. ELM acknowledges support from the National Science Foundation's ID/R Program.

\bibliographystyle{IEEEtran}
\bibliography{references}

\end{document}